\documentclass[conference]{IEEEtran}

\usepackage{cite}
\usepackage{amsmath,amssymb,amsfonts}
\usepackage{graphicx}
\usepackage{textcomp}
\usepackage{xcolor}
\IEEEoverridecommandlockouts

\def\BibTeX{{\rm B\kern-.05em{\sc i\kern-.025em b}\kern-.08em
    T\kern-.1667em\lower.7ex\hbox{E}\kern-.125emX}}

\begin{document}

\title{Learning-Free Hierarchical Joint Estimation of AoA, Pointing Error, Receiver Jitter, and Turbulence in Multi-Aperture FSO Systems
\thanks{This work was supported by the Qatar Research Development and Innovation Council (QRDI) under Grant No. NPRP14C-0909-210008 and by research funding from Hamad Bin Khalifa University under the Thematic Research Grant Program Cycle 3. The statements made herein are solely the responsibility of the authors. The content is solely the responsibility of the authors and does not necessarily represent the official views of QRDI.}
}

\author{
\IEEEauthorblockN{
Mahtab~Dashtebozorgi\IEEEauthorrefmark{1},
Meysam~Ghanbari\IEEEauthorrefmark{2},
Rula~Ammuri\IEEEauthorrefmark{3},
Mazen~Hasna\IEEEauthorrefmark{4},
and~Khalid~Qaraqe\IEEEauthorrefmark{2}
}

\IEEEauthorblockA{
\IEEEauthorrefmark{1}Email: dashtebozorgimahtab@gmail.com\\
}

\IEEEauthorblockA{
\IEEEauthorrefmark{2}College of Science and Engineering, Hamad Bin Khalifa University, Doha, Qatar.\\
Emails: megh89467@hbku.edu.qa, kqaraqe@hbku.edu.qa
}

\IEEEauthorblockA{
\IEEEauthorrefmark{3}Professionals for Smart Technology, Amman, Jordan.\\
Email: rammuri@pst.jo
}

\IEEEauthorblockA{
\IEEEauthorrefmark{4}Department of Electrical Engineering, Qatar University, Doha, Qatar.\\
Email: hasna@qu.edu.qa
}
}

\maketitle
\begin{abstract}
This paper proposes a learning-free hierarchical estimator for jointly recovering the angle of arrival (AoA), transmitter pointing error, receiver-induced jitter, and per-aperture turbulence coefficients in a multi-aperture free-space optical system. The proposed method exploits the distinct spatial signatures contained in quad-photodetector measurements. First, normalized quadrant imbalance ratios provide an approximate closed-form AoA estimate. Next, AoA-compensated lens powers are transformed into a log-linear regression model for non-iterative pointing-error estimation. Finally, receiver jitter and turbulence coefficients are directly reconstructed after compensating for the estimated angular losses. The method requires neither neural-network training nor exhaustive multidimensional search and has computational complexity linear in the number of lenses. Monte Carlo results demonstrate robust AoA estimation under
Gamma--Gamma turbulence, reveal turbulence-induced error floors in
OLS-based pointing and jitter estimation, and show improved
turbulence reconstruction with larger arrays and explicit pointing
compensation.
\end{abstract}

\begin{IEEEkeywords}
Free-space optical communication, angle-of-arrival
estimation, pointing error, atmospheric turbulence, multi-aperture
receiver.
\end{IEEEkeywords}


\section{Introduction}
The growing demand for high-capacity, secure, and flexible connectivity in next-generation space-air-ground networks has increased interest in free-space optical (FSO) communication. FSO technology is considered a promising solution for applications such as terrestrial backhaul, unmanned aerial vehicle links, inter-satellite communication, and quantum communication \cite{Dabiri2026RealTime}, \cite{DabiriEtAl2025Retroreflector}. However, its performance is highly sensitive to atmospheric turbulence, transmitter pointing errors (PEs), and angle-of-arrival (AoA) fluctuations, particularly in mobile and aerial platforms. Several techniques, including aperture averaging, multi-aperture diversity, adaptive optics, and active beam tracking, have been investigated to mitigate these impairments \cite{PaulEtAl2022PulseJamming,GirdherBansal2024RIS,AtaAlouini2023HAPS,SolankiEtAl2025BeamTracking,Ghanbari2026NarrowBeams}. The effectiveness of these techniques nevertheless depends on accurate and timely estimation of the underlying channel and alignment parameters.
Existing studies have proposed maximum-likelihood, expectation-maximization, and moment-based methods for estimating turbulence parameters \cite{ChenHui2021GammaGamma}, \cite{KimYoon2025MomentBased}, while beam-shape-based approaches have been developed for AoA estimation \cite{DreierEtAl2025AoA}. Most available methods, however, estimate turbulence, PE, and AoA separately. In practical FSO links, these impairments occur simultaneously and are multiplicatively coupled in the received optical power, making their direct joint estimation difficult. A recent hierarchical deep-learning framework demonstrated that the distinct spatial signatures provided by a multi-aperture receiver and quad-photodetectors can support joint estimation of these parameters \cite{DabiriEtAl2026Hierarchical}. Nevertheless, learning-based estimators require extensive training data, network design, and possible retraining when the system geometry or channel statistics change.

In this paper, we propose a learning-free hierarchical framework
for jointly estimating AoA, transmitter PE, receiver jitter, and
per-aperture turbulence coefficients. Normalized quad-PD imbalance
ratios first provide an approximate closed-form AoA estimate.
The AoA-compensated lens powers are then converted into a
log-linear regression for non-iterative PE estimation. Finally,
receiver jitter and turbulence are directly reconstructed after
angular compensation. The resulting estimator has
$\mathcal{O}(N_{\mathrm{lens}})$ complexity and requires neither
training nor multidimensional search. Simulations demonstrate robust AoA estimation under GG turbulence,
reveal turbulence-induced error floors in OLS-based pointing and
jitter estimation, and show reduced turbulence-reconstruction NMSE
with larger arrays and explicit pointing compensation.

\section{System Model}

\begin{figure}[!t]
    \centering
    \includegraphics[width=\columnwidth]{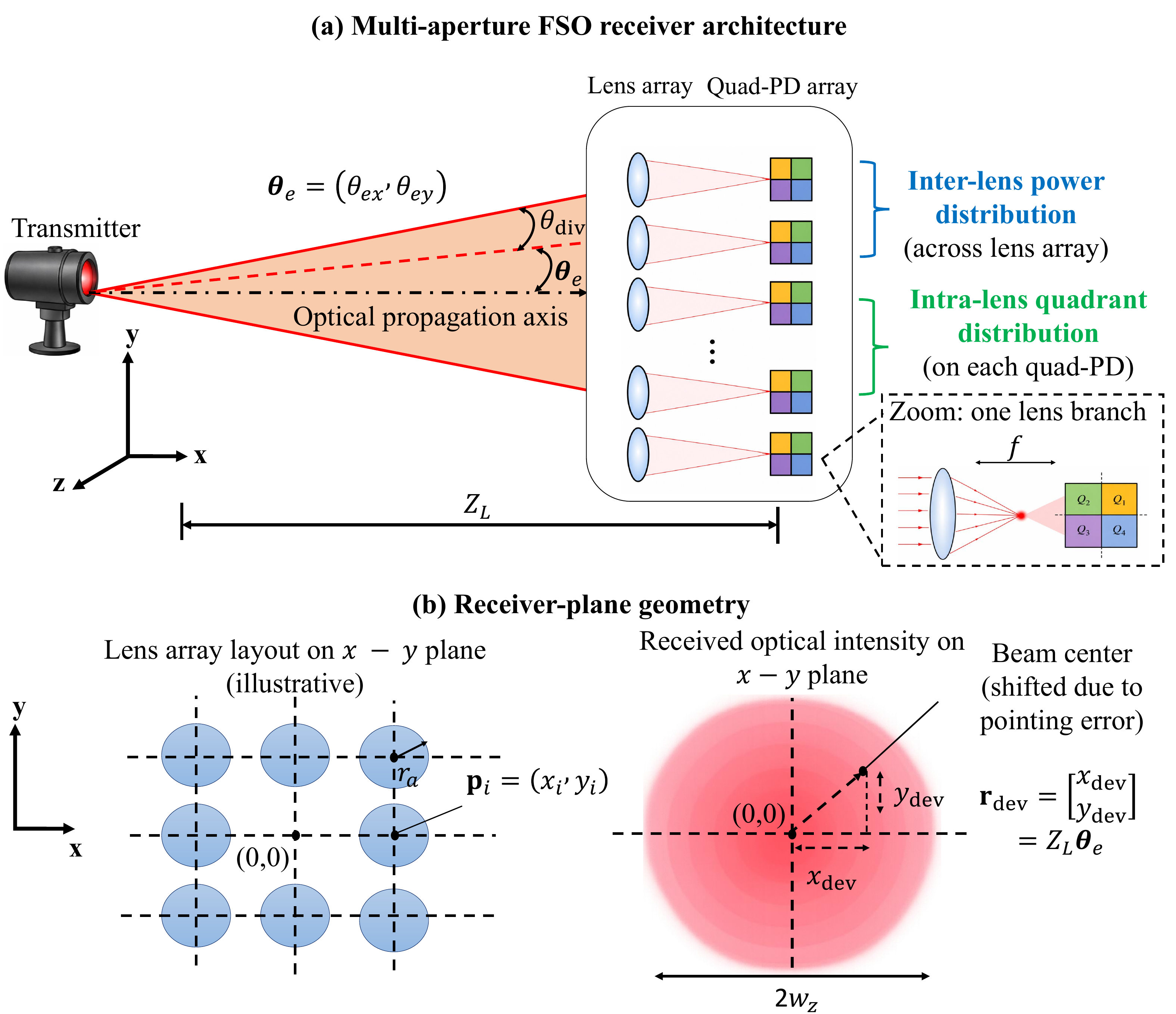}
\caption{Multi-aperture FSO receiver architecture and receiver-plane
geometry, where inter-lens and intra-lens power distributions support
pointing-error/turbulence and AoA estimation, respectively.}
    \label{fig:fig1}
\end{figure}

Fig.~\ref{fig:fig1} shows the considered multi-aperture FSO receiver, in which
an array of small lenses replaces a single large receiving aperture,
and each lens focuses the incident optical beam onto an individual
quad-PD. The inter-lens power distribution supports transmitter
pointing-error and turbulence estimation, whereas the intra-lens
quadrant distribution supports AoA estimation. The lens array lies
in the x–y plane, and optical propagation occurs along the z-axis
over a link distance $Z_{L}$. Transmitter-side tracking imperfections cause the beam axis
to deviate from the ideal line of sight, and the corresponding
transmitter pointing-error vector is defined as
$
\boldsymbol{\theta}_{e}
=
\left(
\theta_{ex},
\theta_{ey}
\right),
$
where $\theta_{e,x}, \theta_{e,y}
\overset{\mathrm{i.i.d.}}{\sim}
\mathcal{N}\left(0,\sigma_{\theta}^{2}\right)$ \cite{DabiriEtAl2026Hierarchical}.
Under the small-angle approximation,
this angular deviation shifts the Gaussian beam center at the
receiver plane according to
\begin{equation}
\mathbf{r}_{\mathrm{dev}}
=
\begin{bmatrix}
x_{\mathrm{dev}}\\
y_{\mathrm{dev}}
\end{bmatrix}
=
Z_{L}\boldsymbol{\theta}_{e}.
\label{eq:beam_deviation_vector}
\end{equation}
Consequently,
$
\mathbf{r}_{\mathrm{dev}}
\sim
\mathcal{N}\left(
\mathbf{0},
\left(Z_L\sigma_{\theta}\right)^2\mathbf{I}_2
\right).
$
The received optical intensity on the receiver plane is modeled
as a circularly symmetric Gaussian beam \cite{DabiriEtAl2026Hierarchical},
\begin{equation}
I(x,y)
=
I_{0}
\exp
\left(
-\frac{
\left(x-x_{\mathrm{dev}}\right)^{2}
+
\left(y-y_{\mathrm{dev}}\right)^{2}
}{
w_{z}^{2}
}
\right),
\label{eq:received_intensity_distribution}
\end{equation}
where $I_{0}$ and $w_{z}$ denote the peak irradiance and beam waist at
distance $Z_{L}$, respectively. The receiver comprises $N_{\mathrm{lens}}$ identical
lenses of aperture radius $r_{a}$, with centers
$
\mathbf{p}_{i}
=
\left(
x_{i},y_{i}
\right)$, $
i\in\{1,\ldots,N_{\mathrm{lens}}\}.
$
With $\mathbf{r}=(x,y)$, the optical power collected by the $i$-th
lens is

\begin{equation}
P_{i}
=
\iint_{\left\lVert \mathbf{r}-\mathbf{p}_{i}\right\rVert\leq r_{a}}
I(\mathbf{r})\,\mathrm{d}\mathbf{r}.
\label{eq:lens_collected_power}
\end{equation}

Under the small-aperture assumption $w_z\gg r_a$, the Gaussian
intensity is approximately constant over each lens aperture;
thus, $P_i \approx I(\mathbf{p}_i)\pi r_a^2$, and substituting \eqref{eq:received_intensity_distribution}
gives

\begin{equation}
P_i
\approx
I_0 \pi r_a^2
\exp\left(
-\frac{
\left(x_i-Z_L\theta_{ex}\right)^2
+
\left(y_i-Z_L\theta_{ey}\right)^2
}{
w_z^2
}
\right).
\label{eq:collected_power_pointing_error}
\end{equation}

Normalizing $P_i$ by the reference Gaussian beam power $I_0\pi w_z^2$ gives the geometric pointing gain of the $i$-th lens as
\begin{equation}
\begin{aligned}
h_{p,i}\left(\theta_{ex},\theta_{ey}\right)
&=
\frac{r_a^2}{w_z^2}
\exp\Bigg(
-\frac{
\left(x_i-Z_L\theta_{ex}\right)^2
}{
w_z^2
}
\\[-1mm]
&\hspace{25mm}
-\frac{
\left(y_i-Z_L\theta_{ey}\right)^2
}{
w_z^2
}
\Bigg).
\end{aligned}
\label{eq:geometric_pointing_gain}
\end{equation}
This spatially varying inter-lens power profile is subsequently
used for log-linear estimation of the transmitter pointing-error
vector. Mobile or vibrating receiver platforms introduce receiver-side angular jitter, denoted by
$
\boldsymbol{\theta}_{r}
=
\left(
\theta_{rx},
\theta_{ry}
\right)
\sim
\mathcal{N}
\left(
\mathbf{0},
\sigma_{r}^{2}\mathbf{I}_{2}
\right),
$
independently of $\boldsymbol{\theta}_{e}$ \cite{DabiriEtAl2026Hierarchical}. The instantaneous angle of arrival is therefore
\begin{equation}
\boldsymbol{\theta}_{\mathrm{AoA}}
=
\left(
\theta_{\mathrm{AoA},x},
\theta_{\mathrm{AoA},y}
\right)
=
\boldsymbol{\theta}_{e}
+
\boldsymbol{\theta}_{r},
\label{eq:instantaneous_aoa}
\end{equation}
and is distributed as
$
\boldsymbol{\theta}_{\mathrm{AoA}}
\sim
\mathcal{N}
\left(
\mathbf{0},
\sigma_{\theta,\mathrm{eff}}^{2}\mathbf{I}_{2}
\right),
$
where
$
\sigma_{\theta,\mathrm{eff}}^{2}
=
\sigma_{\theta}^{2}
+
\sigma_{r}^{2}$. After the $i$-th lens, this AoA displaces the focused spot on its
quad-PD. Under the small-angle approximation, the focal-plane displacement is
\begin{equation}
\mathbf{s}_{i}
=
\begin{bmatrix}
s_{i,x}\\
s_{i,y}
\end{bmatrix}
=
f_{c}\tan\left(\boldsymbol{\theta}_{\mathrm{AoA}}\right)
\approx
f_{c}
\begin{bmatrix}
\theta_{\mathrm{AoA},x}\\
\theta_{\mathrm{AoA},y}
\end{bmatrix},
\label{eq:focal_plane_displacement}
\end{equation}
where $f_c$ is the focal length of each small lens. The focused-spot intensity is approximated by a
two-dimensional Gaussian distribution \cite{DreierEtAl2025AoA},
\begin{equation}
I_{f}^{(i)}\left(\mathbf{r}_{f}\right)
=
\frac{1}{\pi w_{\mathrm{spot}}^{2}}
\exp\left(
-\frac{
\left\lVert
\mathbf{r}_{f}-\mathbf{s}_{i}
\right\rVert^{2}
}{
w_{\mathrm{spot}}^{2}
}
\right),
\label{eq:focal_plane_psf}
\end{equation}
where
$\mathbf{r}_{f}=\begin{bmatrix}x_{f}\\y_{f}\end{bmatrix}$
is the focal-plane coordinate and $w_{\mathrm{spot}}$ is the effective focal-spot radius. Each quad-PD has a square active area of side length dPD,
partitioned into four equal quadrants:
\begin{subequations}
\label{eq:quad_pd_regions}
\begin{align}
\Omega_{i,1}
&=
\left\{
(x_f,y_f):
0\leq x_f\leq \frac{d_{\mathrm{PD}}}{2},
\;
0\leq y_f\leq \frac{d_{\mathrm{PD}}}{2}
\right\},
\label{eq:quad_pd_region_1}
\\
\Omega_{i,2}
&=
\left\{
(x_f,y_f):
-\frac{d_{\mathrm{PD}}}{2}\leq x_f\leq 0,
\;
0\leq y_f\leq \frac{d_{\mathrm{PD}}}{2}
\right\},
\label{eq:quad_pd_region_2}
\\
\Omega_{i,3}
&=
\left\{
(x_f,y_f):
-\frac{d_{\mathrm{PD}}}{2}\leq x_f\leq 0,
\;
-\frac{d_{\mathrm{PD}}}{2}\leq y_f\leq 0
\right\},
\label{eq:quad_pd_region_3}
\\
\Omega_{i,4}
&=
\left\{
(x_f,y_f):
0\leq x_f\leq \frac{d_{\mathrm{PD}}}{2},
\;
-\frac{d_{\mathrm{PD}}}{2}\leq y_f\leq 0
\right\}.
\label{eq:quad_pd_region_4}
\end{align}
\end{subequations}
The AoA-induced gain of quadrant $j$ behind lens $i$ is defined
as the fraction of the focused Gaussian-spot energy within
$\Omega_{i,j}$:

\begin{equation}
\begin{aligned}
h_{\mathrm{AoA},i,j}
\left(
\boldsymbol{\theta}_{\mathrm{AoA}}
\right)
&=
\iint_{\Omega_{i,j}}
\frac{1}{\pi w_{\mathrm{spot}}^{2}}
\\[-1mm]
&\quad\times
\exp\left(
-\frac{
\left\lVert
\mathbf{r}_{f}
-
f_{c}\boldsymbol{\theta}_{\mathrm{AoA}}
\right\rVert^{2}
}{
w_{\mathrm{spot}}^{2}
}
\right)
\mathrm{d}\mathbf{r}_{f}.
\end{aligned}
\label{eq:aoa_induced_quadrant_gain}
\end{equation}

Since the Gaussian PSF in \eqref{eq:focal_plane_psf} is separable along the $x_f$ and $y_f$ directions, its integral over
$[x_L,x_U]\times[y_L,y_U]$ is
\begin{equation}
\begin{aligned}
H_{\mathrm{AoA}}
\left(
x_L,x_U;y_L,y_U
\right)
&=
\left[
\Phi\left(
\frac{x_U-s_x}{\sigma}
\right)
-
\Phi\left(
\frac{x_L-s_x}{\sigma}
\right)
\right]
\\
&\quad\times
\left[
\Phi\left(
\frac{y_U-s_y}{\sigma}
\right)
-
\Phi\left(
\frac{y_L-s_y}{\sigma}
\right)
\right],
\end{aligned}
\label{eq:rectangular_aoa_gain}
\end{equation}
where $\Phi(\cdot)$ is the standard normal cumulative distribution function,
$s_x=f_c\theta_{\mathrm{AoA},x}$,
$s_y=f_c\theta_{\mathrm{AoA},y}$, and
$\sigma=w_{\mathrm{spot}}/\sqrt{2}$.
The four quadrant gains are then obtained by evaluating
$H_{\mathrm{AoA}}(x_L,x_U;y_L,y_U)$ over the limits in
\eqref{eq:quad_pd_region_1}--\eqref{eq:quad_pd_region_4}.
The nonnegative aperture turbulence coefficients
$\{h_{a,i}\}_{i=1}^{N_{\mathrm{lens}}}$ follow Gamma--Gamma (GG)
distributions with small- and large-scale parameters $\alpha$ and
$\beta$ \cite{ChenHui2021GammaGamma}, and are assumed independent
for sufficiently separated apertures
\cite{DabiriEtAl2026Hierarchical}.
The instantaneous photocurrent at quadrant $j$ behind lens $i$ is

\begin{equation}
y_{i,j}[\ell]
=
P_t h_0
h_{p,i}\!\left(\boldsymbol{\theta}_e\right)
h_{\mathrm{AoA},i,j}\!\left(\boldsymbol{\theta}_{\mathrm{AoA}}\right)
h_{a,i}
+
n_{i,j}[\ell],
\label{eq:instantaneous_quad_pd_output}
\end{equation}
where $P_t$, $h_0$, and $
n_{i,j}[\ell]
\sim
\mathcal{N}\!\left(0,\sigma_n^2\right)
$ denote the transmit
optical power, deterministic optical losses, and additive receiver
noise, respectively, while $\ell$ indexes the samples within a channel
coherence block. Assuming constant channel parameters over $L_b$
samples, the block-averaged quad-PD output is

\begin{equation}
\bar{y}_{i,j}
=
\frac{1}{L_b}
\sum_{\ell=1}^{L_b}
y_{i,j}[\ell].
\label{eq:block_averaged_quad_pd_output}
\end{equation}
This yields
\begin{equation}
\bar{y}_{i,j}
=
P_t h_0
h_{p,i}\!\left(\boldsymbol{\theta}_e\right)
h_{\mathrm{AoA},i,j}\!\left(\boldsymbol{\theta}_{\mathrm{AoA}}\right)
h_{a,i}
+
\bar{n}_{i,j},
\label{eq:block_averaged_measurement_model}
\end{equation}
where
$
\bar{n}_{i,j}
\sim
\mathcal{N}
\left(
0,
\frac{\sigma_n^2}{L_b}
\right).
$
Defining the block-averaged noise variance as
$\sigma_{\bar{n}}^{2}\triangleq\sigma_n^2/L_b$ gives
$\bar{n}_{i,j}\sim\mathcal{N}(0,\sigma_{\bar{n}}^{2})$. The unknown parameter set is

\begin{equation}
\boldsymbol{\Theta}
=
\left\{
\boldsymbol{\theta}_{\mathrm{AoA}},
\boldsymbol{\theta}_e,
\left\{h_{a,i}\right\}_{i=1}^{N_{\mathrm{lens}}}
\right\}.
\label{eq:unknown_parameter_set}
\end{equation}

The receiver-induced jitter $\boldsymbol{\theta}_r$ is not an independent unknown
and is reconstructed from the additive AoA model after estimating
$\boldsymbol{\theta}_{\mathrm{AoA}}$ and $\boldsymbol{\theta}_e$. The proposed learning-free hierarchical estimator
recovers $\boldsymbol{\Theta}$ from the block-averaged quad-PD measurements $\left\{\bar{y}_{i,j}\right\}$
without neural-network training or exhaustive multidimensional
search.

\section{Proposed Estimation Method}

\subsection{Estimator Overview}

The block-averaged quad-PD measurements in~\eqref{eq:block_averaged_measurement_model} contain the coupled effects of
transmitter pointing error, AoA-induced focal-plane displacement, and
aperture-dependent turbulence fading. Their direct joint estimation is challenging
because of the multiplicative structure of the measurement model in~\eqref{eq:block_averaged_measurement_model}.
The proposed estimator exploits the distinct signatures of these impairments in the multi-aperture quad-PD measurements. The AoA mainly affects the relative optical-power distribution among the four quadrants of each quad-PD, whereas transmitter pointing error affects the spatial power profile across the lens array. After compensating for AoA and pointing loss, the remaining aperture-dependent power fluctuation is attributed to turbulence fading. Accordingly, the proposed learning-free hierarchical framework decouples the estimation problem into three sequential stages rather than learning the nonlinear mapping from
$\{\bar{y}_{i,j}\}$
to the unknown parameters using neural networks.
First, the AoA vector
$\boldsymbol{\theta}_{\mathrm{AoA}}$
is estimated from normalized quad-PD imbalance ratios, which suppress the common multiplicative effects of transmit power, deterministic loss, pointing loss, and turbulence within each lens. Second, the estimated AoA is used to remove the AoA contribution from the total received power of each lens. The resulting per-lens power sequence is transformed into a log-linear spatial regression problem, from which the transmitter pointing-error vector
$\boldsymbol{\theta}_e$
is estimated by least squares. Third, after estimating both AoA and transmitter pointing error, the receiver jitter and per-aperture turbulence coefficients are directly reconstructed. The overall estimation flow is summarized as
\begin{equation}
\{\bar{y}_{i,j}\}
\longrightarrow
\widehat{\boldsymbol{\theta}}_{\mathrm{AoA}}
\longrightarrow
\widehat{\boldsymbol{\theta}}_e
\longrightarrow
\left\{
\left\{\widehat{h}_{a,i}\right\}_{i=1}^{N_{\mathrm{lens}}},
\widehat{\boldsymbol{\theta}}_r
\right\}.
\label{eq:estimation_flow}
\end{equation}

The hierarchical ordering in \eqref{eq:estimation_flow} is physically motivated.
AoA compensation yields the lens-wise spatial power profile
required for log-linear pointing-error estimation, while removing
both angular effects subsequently enables turbulence reconstruction.
The receiver jitter is obtained as
$\widehat{\boldsymbol{\theta}}_{r}
=\widehat{\boldsymbol{\theta}}_{\mathrm{AoA}}
-\widehat{\boldsymbol{\theta}}_{e}$.

\subsection{Closed-Form AoA Estimation}

The first stage estimates the instantaneous AoA vector
$\boldsymbol{\theta}_{\mathrm{AoA}}$ from the relative power distribution among the four quadrants of each quad-PD. From \eqref{eq:block_averaged_measurement_model}, for a fixed lens $i$, the terms $P_t$, $h_0$, $h_{p,i}(\boldsymbol{\theta}_e)$, and $h_{a,i}$ are common to all four quadrants. Hence, quadrant imbalance ratios suppress these common multiplicative factors and isolate the effect of the AoA-induced focal-plane displacement. The total received quad-PD power at the $i$-th lens is defined as
\begin{equation}
\bar{y}_{i}^{\Sigma}
=
\sum_{j=1}^{4}
\bar{y}_{i,j}.
\label{eq:total_quad_pd_power}
\end{equation}

The normalized horizontal and vertical imbalance ratios are
\begin{equation}
D_{x,i}
=
\frac{
\left(\bar{y}_{i,1}+\bar{y}_{i,4}\right)
-
\left(\bar{y}_{i,2}+\bar{y}_{i,3}\right)
}{
\bar{y}_{i}^{\Sigma}
},
\label{eq:horizontal_imbalance_ratio}
\end{equation}
\begin{equation}
D_{y,i}
=
\frac{
\left(\bar{y}_{i,1}+\bar{y}_{i,2}\right)
-
\left(\bar{y}_{i,3}+\bar{y}_{i,4}\right)
}{
\bar{y}_{i}^{\Sigma}
}.
\label{eq:vertical_imbalance_ratio}
\end{equation}

Here, $D_{x,i}$ measures the normalized power difference between the right and left halves of the quad-PD, whereas $D_{y,i}$ measures that between the upper and lower halves. Since the AoA shifts the focused spot on the focal plane, these ratios provide direct information about the horizontal and vertical components of $\boldsymbol{\theta}_{\mathrm{AoA}}$. 
To avoid additive-noise-induced numerical instability, the measured
imbalance ratios are clipped as
\begin{equation}
\widetilde{D}_{u,i}
=
\min
\left\{
1-\delta,\,
\max\left\{-1+\delta,D_{u,i}\right\}
\right\},
\qquad
u\in\{x,y\},
\label{eq:clipped_imbalance}
\end{equation}
where $0<\delta\ll1$, ensuring that the arguments of
$\operatorname{erf}^{-1}(\cdot)$ remain within its valid domain.
Using the Gaussian PSF model in~\eqref{eq:focal_plane_psf}, the
right--left and upper--lower imbalance ratios are related to the
focal-plane spot displacement. When the quad-PD active area is
sufficiently larger than the focal-spot radius and the displaced spot
remains within the detector support, finite-area truncation becomes
negligible. Under this condition,

\begin{equation}
D_{x,i}
\approx
\operatorname{erf}
\left(
\frac{s_x}{w_{\mathrm{spot}}}
\right),
\label{eq:dx_erf}
\end{equation}
\begin{equation}
D_{y,i}
\approx
\operatorname{erf}
\left(
\frac{s_y}{w_{\mathrm{spot}}}
\right),
\label{eq:vertical_imbalance_erf}
\end{equation}

where the AoA-induced focal-plane displacements follow from
\eqref{eq:focal_plane_displacement} as
$s_x=f_c\theta_{\mathrm{AoA},x}$ and
$s_y=f_c\theta_{\mathrm{AoA},y}$.
Therefore, under the negligible detector-truncation approximation,
the per-lens AoA components are estimated as
\begin{equation}
\widehat{\theta}_{\mathrm{AoA},u,i}
=
\frac{w_{\mathrm{spot}}}{f_c}
\operatorname{erf}^{-1}
\left(
\widetilde{D}_{u,i}
\right),
\qquad
u\in\{x,y\}.
\label{eq:per_lens_aoa_components}
\end{equation}

To reduce the effect of weak and noisy apertures, the per-lens
estimates are combined using $W_i=\overline{y}^{\Sigma}_i$ as
\begin{equation}
\widehat{\theta}_{\mathrm{AoA},u}
=
\frac{
\sum_{i=1}^{N_{\mathrm{lens}}}
W_i\widehat{\theta}_{\mathrm{AoA},u,i}
}{
\sum_{i=1}^{N_{\mathrm{lens}}}W_i
},
\qquad
u\in\{x,y\}.
\label{eq:weighted_aoa_components}
\end{equation}

Thus,
$\widehat{\boldsymbol{\theta}}_{\mathrm{AoA}}
=
(\widehat{\theta}_{\mathrm{AoA},x},
\widehat{\theta}_{\mathrm{AoA},y})$.
The power weights favor stronger lens observations. Hence, this
stage provides a low-complexity approximate closed-form AoA
estimate under negligible detector truncation, without neural-network
training or iterative search.

\subsection{Log-Linear Pointing-Error Estimation}

The second stage estimates the transmitter pointing-error vector
$\boldsymbol{\theta}_e=(\theta_{ex},\theta_{ey})$ from the spatial
power profile across the lens array after AoA compensation.
Using the total quad-PD power $\bar y_i^\Sigma$ defined in
\eqref{eq:total_quad_pd_power}, summing
\eqref{eq:block_averaged_measurement_model} over the four quadrants gives
\begin{equation}
\bar y_i^\Sigma
=
P_t h_0
h_{p,i}\!\left(\boldsymbol{\theta}_e\right)
h_{a,i}
\sum_{j=1}^{4}
h_{\mathrm{AoA},i,j}
\!\left(\boldsymbol{\theta}_{\mathrm{AoA}}\right)
+
\bar n_i^\Sigma,
\label{eq:summed_quad_pd_measurement}
\end{equation}
where
$\bar n_i^\Sigma=\sum_{j=1}^{4}\bar n_{i,j}$.
Using the first-stage AoA estimate, the total AoA-induced gain is

\begin{equation}
\widehat{h}_{\mathrm{AoA},i}^{\mathrm{tot}}
=
\sum_{j=1}^{4}
h_{\mathrm{AoA},i,j}
\left(
\widehat{\boldsymbol{\theta}}_{\mathrm{AoA}}
\right).
\label{eq:estimated_total_aoa_gain}
\end{equation}

The AoA-compensated per-lens power is defined as
\begin{equation}
z_i
=
\frac{
\bar{y}_{i}^{\Sigma}
}{
P_t h_0
\widehat{h}_{\mathrm{AoA},i}^{\mathrm{tot}}
}.
\label{eq:aoa_compensated_lens_power}
\end{equation}
For an accurate AoA estimate, the compensated per-lens power satisfies
$
z_i
\approx
h_{p,i}\!\left(\boldsymbol{\theta}_{e}\right)
h_{a,i}.
$
Using \eqref{eq:geometric_pointing_gain},
\begin{equation}
z_i
\approx
\frac{r_a^2}{w_z^2}
\exp\left(
-\frac{
\left(x_i-Z_L\theta_{ex}\right)^2
+
\left(y_i-Z_L\theta_{ey}\right)^2
}{
w_z^2
}
\right)
h_{a,i}.
\label{eq:compensated_power_expanded}
\end{equation}

To avoid numerical instability before the logarithmic transformation,
the compensated power is lower-bounded as
$z_i^{+}=\max\{z_i,\epsilon_z\}$, where $0<\epsilon_z\ll1$.
Using $z_i^{+}$ in the log-linear regression,
\begin{equation}
\ln z_i^{+}
\approx
\ln\left(
\frac{r_a^2}{w_z^2}
\right)
-
\frac{
\left(x_i-Z_L\theta_{ex}\right)^2
+
\left(y_i-Z_L\theta_{ey}\right)^2
}{
w_z^2
}
+
\ln h_{a,i}.
\label{eq:log_compensated_power}
\end{equation}
Expanding the quadratic terms gives
\begin{equation}
\begin{aligned}
\ln z_i^{+}
&\approx
\ln\left(
\frac{r_a^2}{w_z^2}
\right)
-
\frac{x_i^2+y_i^2}{w_z^2}
+
\frac{2Z_L}{w_z^2}
\left(
x_i\theta_{ex}
+
y_i\theta_{ey}
\right)
\\
&\quad
-
\frac{Z_L^2}{w_z^2}
\left(
\theta_{ex}^2+\theta_{ey}^2
\right)
+
\ln h_{a,i}.
\end{aligned}
\label{eq:expanded_log_compensated_power}
\end{equation}

Now define the transformed observation

\begin{equation}
q_i
=
\ln z_i^{+}
+
\frac{x_i^2+y_i^2}{w_z^2}.
\label{eq:transformed_observation}
\end{equation}

Then, \eqref{eq:expanded_log_compensated_power} can be written as the log-linear spatial regression
model $q_i=a+b_xx_i+b_yy_i+\epsilon_i$, where
$b_x=2Z_L\theta_{ex}/w_z^2$ and
$b_y=2Z_L\theta_{ey}/w_z^2$.
The intercept $a$ collects spatially invariant terms,
including $\ln(r_a^2/w_z^2)$ and
$-Z_L^2(\theta_{ex}^2+\theta_{ey}^2)/w_z^2$, whereas
$\epsilon_i$ represents the spatially unstructured residual due
mainly to $\ln h_{a,i}$ and noise. Since the turbulence coefficients are independent of the lens
coordinates and identically distributed across apertures, this
scalar model can be written in matrix form as
\begin{equation}
\mathbf{q}
=
\mathbf{X}\mathbf{b}
+
\boldsymbol{\epsilon},
\label{eq:matrix_log_linear_model}
\end{equation}
where
$\mathbf q=[q_1,\ldots,q_{N_{\mathrm{lens}}}]^T$,
$\mathbf b=[a,b_x,b_y]^T$, and the $i$-th row of
$\mathbf X$ is $[1,x_i,y_i]$.
The regression matrix $\mathbf{X}$ must have full column rank,
which requires at least three non-collinear lens centers. Under this
condition, the least-squares estimate of $\mathbf{b}$ is
\begin{equation}
\widehat{\mathbf{b}}
=
\left(
\mathbf{X}^{T}\mathbf{X}
\right)^{-1}
\mathbf{X}^{T}\mathbf{q}.
\label{eq:least_squares_regression}
\end{equation}

\begin{table*}[t]
\centering
\caption{Angular-estimation RMSE under deterministic and
Gamma--Gamma aperture gains. All values are in $\mu\mathrm{rad}$.}
\label{tab:angular_rmse_turbulence}
\begin{tabular}{c c c c c c}
\hline
Gain model & SNR & Estimator & AoA & Pointing & Jitter \\
\hline
$h_{a,i}=1$
& 20 dB
& Proposed OLS
& 4.41
& 2.75
& 5.17
\\

$h_{a,i}=1$
& 40 dB
& Proposed OLS
& 0.44
& 0.27
& 0.52
\\

GG, $\alpha=\beta=10$
& 20 dB
& Proposed OLS
& 4.88
& 19.79
& 20.35
\\

GG, $\alpha=\beta=10$
& 40 dB
& Proposed OLS
& 0.48
& 17.93
& 17.94
\\

GG, $\alpha=\beta=10$
& Noiseless
& Proposed OLS
& $<10^{-9}$
& 17.92
& 17.92
\\

GG, $\alpha=\beta=10$
& Noiseless
& Oracle turb.\ comp.
& $<10^{-9}$
& $<10^{-9}$
& $<10^{-9}$
\\
\hline
\end{tabular}
\end{table*}

Using additional lenses improves robustness against turbulence
fluctuations and receiver noise.
The transmitter pointing-error components are recovered as
\begin{align}
\widehat{\theta}_{ex}
&=
\frac{w_z^2}{2Z_L}
\widehat{b}_x,
\label{eq:estimated_pointing_error_x}
\\
\widehat{\theta}_{ey}
&=
\frac{w_z^2}{2Z_L}
\widehat{b}_y,
\label{eq:estimated_pointing_error_y}
\end{align}
yielding
$\widehat{\boldsymbol{\theta}}_e
=(\widehat{\theta}_{ex},\widehat{\theta}_{ey})$.
The nonlinear pointing-error problem is therefore reduced to a
non-iterative linear regression. The angular-estimation results in Fig.~2 and Table~I use the OLS
estimator in \eqref{eq:least_squares_regression}. For Fig.~3, the received-power-weighted estimate
is
$\widehat{\mathbf b}_{\mathrm{WLS}}
=(\mathbf X^{T}\mathbf W\mathbf X)^{-1}
\mathbf X^{T}\mathbf W\mathbf q$,
where $\mathbf W=\operatorname{diag}(w_1,\ldots,w_{N_{\mathrm{lens}}})$
and $w_i=z_i^{+}/\max_k z_k^{+}$, reducing the influence of weakly
illuminated apertures.

\subsection{Turbulence and Receiver-Jitter Reconstruction}

\begin{figure}[!t]
    \centering
    \includegraphics[width=\columnwidth]{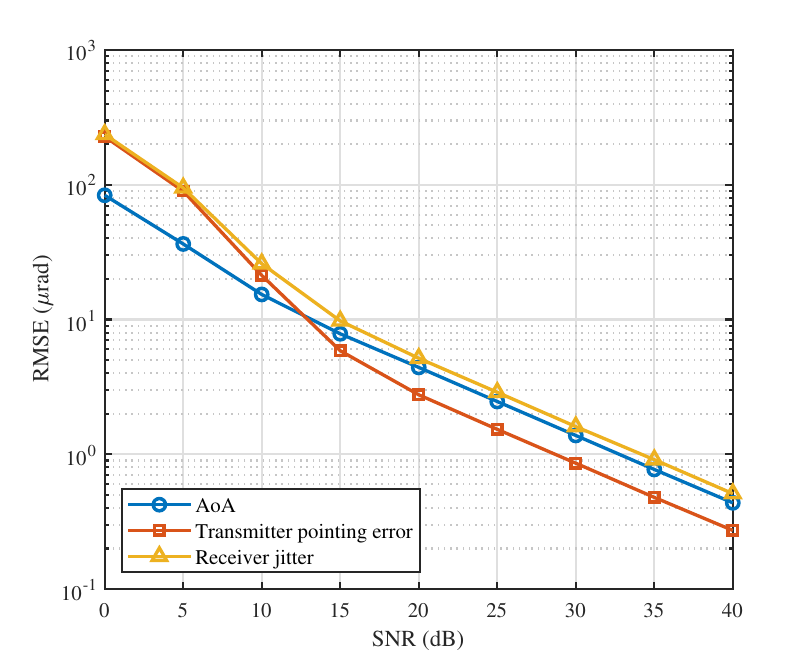}
\caption{Angular-estimation RMSE versus SNR for the proposed OLS
estimator using a $6\times6$ lens array with $h_{a,i}=1$, isolating
receiver-noise effects.}
    \label{fig:fig2new}
\end{figure}

The final stage reconstructs the per-aperture turbulence coefficients
and receiver-induced jitter. Using the pointing-gain model in \eqref{eq:geometric_pointing_gain}, the estimated
pointing gain of the $i$-th lens is obtained by evaluating it at the
estimated transmitter pointing-error vector:
$
\widehat{h}_{p,i}
=
h_{p,i}
\left(
\widehat{\theta}_{ex},
\widehat{\theta}_{ey}
\right).
$
Hence, using \eqref{eq:aoa_compensated_lens_power}, the turbulence coefficient
is equivalently reconstructed as
\begin{equation}
\widehat{h}_{a,i}
=
\frac{z_i}{\widehat{h}_{p,i}}
=
\frac{
\overline{y}^{\Sigma}_{i}
}{
P_t h_0
\widehat{h}^{\mathrm{tot}}_{\mathrm{AoA},i}
\widehat{h}_{p,i}
},
\qquad
i=1,\ldots,N_{\mathrm{lens}}.
\label{eq:equivalent_turbulence_estimate}
\end{equation}

Thus, the turbulence coefficient is obtained after compensating the
deterministic loss, AoA-induced collection loss, and pointing loss. To enforce nonnegativity and prevent unstable division in weakly
illuminated branches, define
$D_i=P_t h_0\widehat{h}_{\mathrm{AoA},i}^{\mathrm{tot}}
\widehat{h}_{p,i}$.
The regularized denominator is
\begin{equation}
D_i^{+}
=
\max\left\{
D_i,
\eta_D \max_k D_k
\right\},
\qquad
0<\eta_D\ll 1,
\label{eq:thresholded_turbulence_denominator}
\end{equation}
and the practical turbulence estimate is
\begin{equation}
\widehat{h}_{a,i}
=
\max\left\{
0,
\frac{\bar y_i^{\Sigma}}{D_i^{+}}
\right\}.
\label{eq:practical_turbulence_estimate}
\end{equation}
This regularization limits noise amplification in weakly illuminated
branches. Using the additive AoA model in
\eqref{eq:instantaneous_aoa}, the receiver-jitter estimate is
$
\widehat{\boldsymbol{\theta}}_{r}
=
\widehat{\boldsymbol{\theta}}_{\mathrm{AoA}}
-
\widehat{\boldsymbol{\theta}}_{e}.
$
Therefore, the final augmented reconstructed parameter set is
\begin{equation}
\widehat{\boldsymbol{\Theta}}_{\mathrm{aug}}
=
\left\{
\widehat{\boldsymbol{\theta}}_{\mathrm{AoA}},
\widehat{\boldsymbol{\theta}}_{e},
\widehat{\boldsymbol{\theta}}_{r},
\left\{
\widehat{h}_{a,i}
\right\}_{i=1}^{N_{\mathrm{lens}}}
\right\}.
\label{eq:final_reconstructed_parameter_set}
\end{equation}
The complexity scales as $\mathcal{O}(N_{\mathrm{lens}})$, since the
lens-wise operations scale linearly with $N_{\mathrm{lens}}$ and the least-squares stage only requires the inversion of a fixed-size $3\times3$
matrix.

\section{Simulation Results}

Monte Carlo simulations are used to evaluate the proposed
hierarchical estimator. Unless otherwise stated,
$Z_L=1000$ m, $w_z=0.15$ m, $r_a=0.01$ m,
$f_c=0.05$ m, $w_{\mathrm{spot}}=20~\mu\mathrm{m}$,
$d_{\mathrm{PD}}=200~\mu\mathrm{m}$, $P_t=h_0=1$,
$\sigma_\theta=20~\mu\mathrm{rad}$,
$\sigma_r=10~\mu\mathrm{rad}$, and
$N_{\mathrm{MC}}=8000$. For Fig.~\ref{fig:fig2new} and Table~I, a \(6\times6\) lens array is used, corresponding to \(N_{\mathrm{lens}}=36\), with \(d_{\mathrm{lens}}=0.04~\mathrm{m}\). Fig.~2 and Table~I use OLS, whereas Fig.~3 uses the
received-power-weighted variant described in Section~III-C. The numerical constants used for imbalance-ratio clipping, log-domain stabilization, and denominator regularization are set to \(\delta=10^{-6}\), \(\epsilon_z=10^{-12}\), and \(\eta_D=10^{-6}\), respectively. For each realization, the SNR is defined from the average squared
noise-free quad-PD output over all lenses and quadrants. The
noise-free block-averaged measurement is
\begin{equation}
\bar{y}_{i,j}^{\mathrm{clean}}
=
P_t h_0
h_{p,i}\!\left(\boldsymbol{\theta}_e\right)
h_{\mathrm{AoA},i,j}\!\left(\boldsymbol{\theta}_{\mathrm{AoA}}\right)
h_{a,i},
\label{eq:clean_measurement}
\end{equation}
and the corresponding average signal power is
\begin{equation}
P_{\mathrm{sig}}
=
\frac{1}{4N_{\mathrm{lens}}}
\sum_{i=1}^{N_{\mathrm{lens}}}
\sum_{j=1}^{4}
\left(
\bar{y}_{i,j}^{\mathrm{clean}}
\right)^2.
\label{eq:average_signal_power}
\end{equation}

For a prescribed $\mathrm{SNR}_{\mathrm{dB}}$, the block-averaged
noise variance is set as
$\sigma_{\bar{n}}^{2}=\frac{P_{\mathrm{sig}}}{10^{\mathrm{SNR}_{\mathrm{dB}}/10}}$.
Thus, $P_{\mathrm{sig}}/\sigma_{\bar n}^{2}$ equals the
prescribed SNR in every realization, providing realization-wise
normalization by the average noise-free signal power; since
block-averaged measurements are simulated directly, $L_b$ and
$\sigma_n^{2}$ need not be specified separately. For the noiseless rows in Table~I, the additive noise is set to
zero. In the oracle benchmark, the true aperture turbulence
coefficients are removed before the log-linear regression to isolate
the effect of unknown turbulence on pointing-error estimation. The angular-estimation root-mean-square error (RMSE) is defined as
\begin{equation}
\mathrm{RMSE}_{\boldsymbol{\theta}}
=
\sqrt{
\frac{1}{N_{\mathrm{MC}}}
\sum_{m=1}^{N_{\mathrm{MC}}}
\left\|
\widehat{\boldsymbol{\theta}}^{(m)}
-
\boldsymbol{\theta}^{(m)}
\right\|_2^2
}.
\label{eq:angular_rmse}
\end{equation}

Here, $\boldsymbol{\theta}$ denotes
$\boldsymbol{\theta}_{\mathrm{AoA}}$,
$\boldsymbol{\theta}_{e}$, or
$\boldsymbol{\theta}_{r}$; hence, all reported values are
two-dimensional vector RMSEs.
For Fig.~2, $h_{a,i}=1$ is used to isolate receiver-noise effects.
The AoA, pointing-error, and receiver-jitter RMSEs all decrease
with SNR. 
The direct jitter reconstruction inherits the pointing-estimation
error and therefore exhibits the same GG-induced error floor.
Table~I shows that the normalized quadrant-imbalance ratios largely
cancel the common turbulence coefficient within each quad-PD,
keeping AoA estimation robust under GG turbulence. In contrast,
the term $\ln h_{a,i}$ in \eqref{eq:expanded_log_compensated_power} acts as an SNR-independent
disturbance in the pointing-error regression, producing similar
pointing and jitter floors at 20 and 40 dB. The noiseless OLS row
confirms that these floors originate from unknown aperture
turbulence rather than additive noise, while the oracle benchmark removes this coupling and reduces all
angular RMSEs below $10^{-9}~\mu\mathrm{rad}$.
The turbulence-reconstruction normalized mean-square error (NMSE) is defined as
\begin{equation}
\mathrm{NMSE}_{h_a}
=
\frac{
\sum_{m=1}^{N_{\mathrm{MC}}}
\left\|
\widehat{\mathbf h}_a^{(m)}
-
\mathbf h_a^{(m)}
\right\|_2^2
}{
\sum_{m=1}^{N_{\mathrm{MC}}}
\left\|
\mathbf h_a^{(m)}
\right\|_2^2
}.
\label{eq:turbulence_nmse}
\end{equation}

\begin{figure}[!t]
    \centering
    \includegraphics[width=\columnwidth]{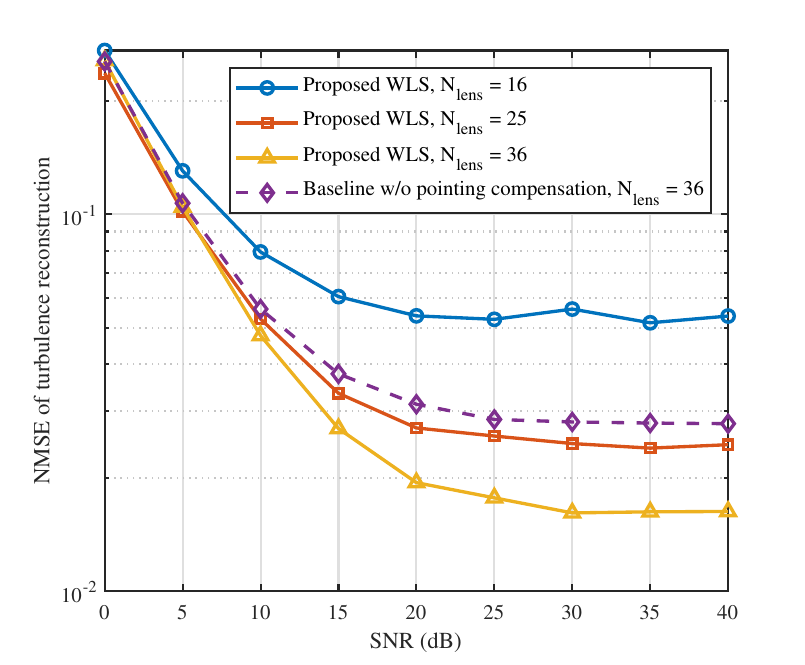}
\caption{Turbulence-reconstruction NMSE versus SNR for different
square lens arrays under GG turbulence with $\alpha=\beta=10$.
The proposed curves use received-power-weighted regression, whereas
the baseline omits pointing-error compensation.}
    \label{fig:fig3}
\end{figure}

For Fig.~3, GG turbulence with $\alpha=\beta=10$ is considered
for $4\times4$, $5\times5$, and $6\times6$ arrays with
$d_{\mathrm{lens}}=0.04$ m. The baseline uses $N_{\mathrm{lens}}=36$ and sets
$\widehat{\boldsymbol{\theta}}_{e}=\mathbf{0}$ when evaluating
the pointing gain in (5), while retaining AoA compensation and the
same denominator regularization. At low SNR, receiver noise dominates and the curves
are close. At moderate and high SNR, increasing the array size
reduces the NMSE, and the $6\times6$ proposed method achieves a
lower error floor than the no-compensation baseline. This confirms
that larger spatial apertures and pointing compensation improve
turbulence reconstruction.

\section{Conclusion}
This paper presented a learning-free hierarchical framework for jointly estimating AoA, transmitter pointing error, receiver jitter, and per-aperture turbulence in multi-aperture FSO systems. By combining closed-form quad-PD imbalance processing, log-linear spatial regression, and direct parameter reconstruction, the proposed method converts a coupled nonlinear estimation problem into three low-complexity stages. Simulation results showed robust AoA estimation under GG turbulence,
while independent aperture turbulence produced error floors in the
OLS-based pointing-error and receiver-jitter estimates. For
turbulence reconstruction, larger arrays and explicit pointing-error
compensation reduced the WLS-based NMSE. The proposed framework
therefore provides a low-complexity and interpretable alternative
to training-based methods while clearly exposing its
turbulence-induced angular-estimation limitation.

\bibliographystyle{IEEEtran}
\bibliography{myref}

\end{document}